\documentclass[conference]{IEEEtran}
\IEEEoverridecommandlockouts
\usepackage{cite}
\usepackage{amsmath,amssymb,amsfonts}
\usepackage{algorithmic}
\usepackage{graphicx}
\usepackage{textcomp}
\usepackage{xcolor}
\usepackage{booktabs}
\usepackage[normalem]{ulem}
\usepackage{multirow}
\usepackage{pifont}
\usepackage{float}
\usepackage{booktabs}
\usepackage[acronym]{glossaries}
\usepackage{subfig}
\usepackage{float}

\useunder{\uline}{\ul}{}

\def\BibTeX{{\rm B\kern-.05em{\sc i\kern-.025em b}\kern-.08em
    T\kern-.1667em\lower.7ex\hbox{E}\kern-.125emX}}
\begin{document}

\title{Intrusion Detection using Machine Learning Techniques: An Experimental Comparison}

\author{\IEEEauthorblockN{Kathryn-Ann Tait$^{1}$, Jan Sher Khan$^{2}$, Fehaid Alqahtani$^{3}$, Awais Aziz Shah$^{4}$, Fadia Ali Khan$^{5}$, Mujeeb Ur Rehman$^{5}$, \\Wadii Boulila$^{6, 7}$  and Jawad Ahmad$^{1}$}
		\IEEEauthorblockA{$^{1}$School of Computing, Edinburgh Napier University, United Kingdom. \\
			$^{2}$Department of Electrical and Electronics, University of Gaziantep, 27310 Gaziantep, Turkey. \\
				$^{3}$Department of Computer Science, King Fahad Naval Academy, Al Jubail, Saudi Arabia.\\
			$^{4}$Department of Electrical and Information Engineering (DEI),
			Politecnico di Bari, Italy. \\	
			 $^{5}$Department of Electrical Engineering, Riphah International University, Islamabad, Pakistan. \\
			    $^{6}$RIADI Laboratory, University of Manouba, Manouba 2010, Tunisia.\\
			 $^{7}$College of Computer Science and Engineering, Taibah University, Medina 42353, Saudi Arabia.
			}
	}
	\IEEEoverridecommandlockouts
\IEEEpubid{\makebox[\columnwidth]{978-1-6654-1224-7/21/\$31.00~\copyright2021 IEEE \hfill} \hspace{\columnsep}\makebox[\columnwidth]{ }}
\maketitle
\IEEEpubidadjcol

\begin{abstract}
	Due to an exponential increase in the number of cyber-attacks, the need for improved Intrusion Detection Systems (IDS) is apparent than ever. In this regard, Machine Learning (ML) techniques are playing a pivotal role in the early classification of the attacks in case of intrusion detection within the system. However, due to the large number of algorithms available, the selection of the right method is a challenging task. To resolve this issue, this paper analyses some of the current state of the art intrusion detection methods and discusses their pros and cons. Further, a review of different ML methods is carried out with four methods showing to be the most suitable one for classifying attacks. Several algorithms are selected and investigated to evaluate the performance of IDS. These IDS classifies binary and multiclass attacks in terms of detecting whether or not the traffic has been considered as benign or an attack. The experimental results demonstrate that binary classification has greater consistency in their accuracy results which ranged from 0.9938 to 0.9977, while multiclass ranges from 0.9294 to 0.9983. However, it has been also observed that multiclass provides the best results with the algorithm k-Nearest neighbor giving an accuracy score of 0.9983 while the binary classification highest score is 0.9977 from Random Forest. The experimental results demonstrate that multiclass classification produces better performance in terms of intrusion detection by specifically differentiating between the attacks and allowing a more targeted response to an attack.
\end{abstract}


\section{\textbf{Introduction} }
Due to the rapid advancements in Internet, there has been an increasing concern over the number of cyber attacks over the past few years \cite{bhuyan2013network}. In the United Kingdom alone last year, 2019, around 32$\%$ of businesses and 22$\%$ of charities reported having experienced a cyber breach or attack  \cite{finnerty2019cyber}. Intrusion Detection System (IDS) is a technique to identify such kind of attacks. In spite of remarkable success of the introduced methods of Intrusion Detection, there has been an increasing concern over improving the existing methods or introducing new ones \cite{bhuyan2013network,huma2021hybrid}. 

IDS has been in use for a number of years with their objective being to scan network traffic and identify any malicious activities or threats in real-time \cite{denning1987intrusion}. As a security system, like a firewall, an IDS has the security purpose of protecting confidentiality, integrity, and availability, which are the main target of a potential attacker to break \cite{liu2014intrusion}. To determine how well an IDS performs there are certain characteristics set out that a good IDS should possess. These include: accuracy, minimal overhead, good performance, performance, and completeness. 

IDS consists of two kind of attack detection methods i.e., anomaly-based detection and Signature-based (or Misuse detection) detecion. The former is used to compare the behavioral changes of the system in order to discover abnormalities in action or activities of the system, for example, if there is an excessive amount of user requests to the database as compared to the historical data, the system will issue an anomaly detection alarm. This of course is also a disadvantage as not every user should be considered the same. Nor would everyday be the same so creating a dynamic query limit would be more beneficial \cite{chio2018machine}. Another disadvantage can be the adaptation of anomaly detecion in a dynamic environment where the user demands changes with time and not comparable with historic data \cite{liu2014intrusion}. So the conclusion is that anomaly detection is susceptible to false positives \cite{chio2018machine}. In the later method of IDS, information known as fingerprints or signatures about previous successful attacks are stored and rules are written within the IDS. These rules are then used to compare against new packets that enter the network \cite{liu2014intrusion}. The main disadvantage with this type of method is that it is limited to what its existing knowledge and is therefore ineffective when detecting new kind of attacks. This also means that the IDS needs to be constantly updated with the knowledge base of new attacks \cite{ali2020network, shaukat2020intrusion,khan2021iot}. \\

Another factor worth mentioning here is the type of IDS i.e., Host-based IDS (HIDS) and Network-based IDS (NIDS). HIDS monitors an individual device on the network against the user activity such as log files and installations \cite{tidjon2019intrusion,liu2014intrusion}. HIDS are highly successful at indicating whether or not an attack has taken place and allows for individual user activity to be monitored. However, it has been shown that is not suitable against major isolated attacks \cite{liu2014intrusion}. On the other side, NIDS are placed near the firewall in order to monitor the whole network and connected devices for ingress/egress traffic \cite{tidjon2019intrusion}. The packets each have their content analysed for protocols that indicate malicious activity. One of the main advantages of NIDS is that the response time is faster and therefore the threat can be dealt with more effectively. Moreover, NIDS are typically stand-alone devices which does not affect the power consumption of the devices on the network as compared to HIDS. However, a significant draw back with NIDS is their encrypted packets mechanism. Due to the packet being encrypted this means the IDS cannot analyse the data within the packet and therefore identify a risk of attack \cite{liu2014intrusion}. 

The authors in \cite{shenfield2018intelligent}, identify that NIDS are typically signature based. From the above section it can be seen that this type of learning is also not optimal when dealing with unknown threats and are prone to a high false positive rate meaning that NIDS are also susceptible to high false positive rates. This means that the system flags up more activities as malicious than it should. While it is known that this is better than having a high false negative rate, it does create an additional workload for administrators as they have to manually check each occurrence.  In order to resolve the aforementioned issues, Machine Learning (ML) methods can be merged with IDS using Artificial Intelligence (AI) models which will empower the systems with the capability to use its data to educate themselves and enhance themselves without being programmed to do so \cite{rege2018machine, latif2020novel, chio2018machine, Keshari2020, Sober2020, shafique2021detecting,amin2017accelerated}.\\

In terms of security, there are already a number of areas which utilize ML in order to protect users and networks against potential attack. A common area where a vast majority of users have, mostly unknown, interaction with ML is through the use of spam detection filters within email. This is typically used as an example of pattern recognition due to the vast number of features, or characteristics, that can be associated with spam \cite{chio2018machine}. To a large majority of users, spam is a nuisance more than anything but unfortunately, they can have a much more sinister meaning such as, identity theft through tricking a user to click on a link. Issues with spam emails have been around for many years and the numbers of spam mail being received by users has vastly reduced with the introduction of ML into the spam detection. This can be seen with Gmail \cite{rege2018machine}. From 2015, Google claimed to block 99.9$\%$ of spam mail after introducing Artificial Neural Network (ANN) to their spam filtering and later announced to have blocked around a 100 million spam emails every day after introducing their new open source ML framework TensorFlow \cite{Wiggers2019}.

In order to improve the current IDS by utilizing the state of the art ML techniques this work aims to evaluate existing ML algorithms and how these could be utilized in improving the current IDS against evolving intrusion attacks.


This work takes into account six different ML algorithms (five supervised and one unsupervised) and evaluates its performance against a number of metrics. For further insights, each algorithm has been trained and tested in terms of binary and multiclass classification. To gain a further insight of where each algorithm provides the best performance, herein the algorithms have been trained and tested in terms of binary and multiclass classification. The results from this study will allow a better overview of these available algorithms for their adaptability in a specific scenario to achieve an optimal performance whether it is a binary or multiclass classification. Additionally, this paper provides a strong basis for further research into the integration of ML with intrusion detection and to present findings which can aid when selecting an appropriate ML algorithm in for the requirements of the IDS.

The rest of the paper has been organized as follows: Section II provides an overview of the steps performed during the classification and processing of the datasets. The description of the obtained results against each metrics is given in Section III ad the comparison of different ML algorithms is given in Section IV. Finally, Section V concludes the paper.

\section{\textbf{Methodology}}
This section provides an overview of the selection and processing of the datasets, the tools used for the feature extraction from the data.

From the two original datasets (UNSW-NB 15 and CICIDS2017), it was noted that there is a vast amount of data present and this would not be suitable for the requirements of this research. To reduce the number of features present within the datasets, a number of data preparation techniques have to be carried out to make the datasets more useable. This section discusses the use of the software platform Weka and how the software was used for the preparation of the data before it is suitable for utilization by the machine learning algorithms. Additonally, it also discusses how manual preparation of the data has been carried out to prepare the implantation of NN and the use of MATLAB to train and test the data.

\subsection{Feature Selection and Data Preprocessing}
Weka software is be used for the data pre-processing abilities and in particular for a specific attribute evaluator. The evaluator used for the selection in our tests is known as "info gain attribute evaluation" which, as mentioned earlier, provides each attribute a ranking based on how much information it provides in relation to the class or target value which in this case is a feature Called “Label”. The attribute evaluator is vital in the process of reducing the number of features present within the datasets. Therefore, the decision is made to reduce the features to 10 for the both datasets, not including the “Label”. This decision would not only remove redundant data that would not have a great impact on the ML algorithm, but also it would reduce the computational power required for the ML algorithms as they would not be processing as much data.\\

After the attribute evaluate was performed, the output displayed the most influential attributes within that dataset in terms of defining the “Label” feature of whether or not an attack occurred. The top ten of these attributes are extracted and later saved in a separate Comma-Separated Value (CSV) file. Then, the CSV files has been uploaded into MATLAB, to perform the splitting of the datasets for training and testing the data. Dataset were spitted into 70$\%$ for training and the remaining 30$\%$ for testing. The data  points are randomly selected to be included in the 70$\%$ for training. The advantages of this includes ensuring a mixture of data points to be used in the training to make the training stage more reliable.

\subsubsection{Neural Networks Preparation}
During this step, a manual separation of the data has been carried out to make the datasets suitable for NN implementation. For the UNSW-NB 15 dataset, this involved extracting the “Label” category from the main dataset and placing it into an individual CSV file. It is done because the NN implementation requires the data targets to be uploaded separately. The following stage convertes all the BENIGN labels to 0 and the ATTACK labels to 1 since the NN requires the target to be 0 or 1. The requirement of 0 or 1 for the preparation of the multiclass dataset, CICIDS2017, creates more of a challenge. To this end, the targets were again extracted into a separate CSV file to allow them to be uploaded individually. The MATLAB (n.d) page states that in order for multiclass classification to be carried out successfully within NN, the target associated with each attribute will have an x-element class vector, which in this case was a 5-element class vector. This means that for each attribute the target could be one of a possible five elements, these elements being BENIGN or one of the four attacks. This was implemented by placing a 1 in the respective element column and a 0 in the other columns. There was no requirement to split this data into training and testing data as the NN application within MATLAB carried out this task out manually, with a 70$\%$, 15$\%$, 15$\%$ split for training, testing, and validation respectively.

\subsection{Machine Learning}
The machine learning aspect of this project has been carried out in MATLAB. The MATLAB software has the ability to execute ML algorithms through different types of learners. For this work, the classification learner and deep learning toolboxes have been utilised. The classification Learner tab has been used in carrying out the supervised learning which involves Logistic Regression (LG), Support Vector Machine (SVM), k-Nearest Neighbour (KNN), and k-Means. For the implementation of NN, the deep learning toolbox has been used. The unsupervised aspect involved reusing Weka to implement k-means clustering. \\

The majority of supervised learning models have been implemented in MATLAB using the classification learner toolbox. Using the toolbox made the implementation of the algorithms simpler and more reliable as the code has already been proven to work and be reliable due to the number of individuals who have used the software platform. Due to some algorithms not being suitable for both the datasets, LG has been implemented for the binary classification datasets while KNN has been implemented for the multiclass classification. For SVM and KNN, multiple variations of the algorithm have been implemented in order to get an overall better understanding and evaluation of the algorithm’s performance. The algorithms have been chosen randomly but kept same for both binary and multiclass classification.\\

Once the models have been trained, they are exported to the MATLAB workspace where each has been tested using the test data created during the data preparation stage. The following command has been used to carry out the testing:\\
$$yfit<variableName> =$$ $$<trainedModelName>.predictFcn(nameOfTestDataset);$$\\

Then, the produced output, known as $$yfit<variableName>$$ contained the predicted labels for the test data. In order the evaluate how well the model has been trained, the evaluation metrics (discussed later in Section III) have been used. 

For the implementation of the NN model, which was also carried out within MATLAB, the deep learning toolbox has been used. After the data has prepared for this model, it inputs into the neural net pattern recognition application. For the implementation of the NN, 10 neurons have been used for both cases to build a hidden layer. The number of neurons could be changed in future studies to further enhance the algorithm to make it more reliable for accurate predictions. In order to evaluate the performance of the NN, this study considers the confusion matrix which provides the overall view of how the algorithm performs.\\

The unsupervised, k-Means clustering, algorithm has been implemented through Weka tool. The main reason the selection of this tool is that it is simple to use for implementing algorithms. After the datasets had been loaded into Weka, the k-Means clustering algorithm has been selected along the number of clusters (k) being respective to the number of classes in each type of dataset. For example, the binary classification consisted of two clusters while the multiclass dataset consisted of 6 clusters. Due to the nature of Weka and how its produces the output data for the evaluation metrics for clustering, the incorrectly clustered instances will be considered to give an indication of how well the algorithm performed.

\section{\textbf{Results}} 
This section discuss the results collected from the binary and multiclass classification performed in this work. Moreover, a brief evaluation of the results are given with a comparison of the two algorithms presented at the end of the section. The evaluation metrics used to evaluate the performance of the ML algorithms are primarily made up of four separate factors i.e, True Positive (TP), False Negative (FN), False Positive (FP), and True Negative (TN). A definition of each of these can be seen in the Table \ref{tab:E_Matric}. 

\begin{table}
	\center
	\setlength{\tabcolsep}{1em}
	\renewcommand{\arraystretch}{1.5}
	\caption{The factors used for the evaluation metrics.}
	\label{tab:E_Matric}       
	\begin{tabular}{llll} 
		\hline
		True Class $\rightarrow$ & True Positive & True Negative  \\                
		Predicted Class  $\downarrow$  &    &   \\
		\hline
		Predicted Positive & TP & FP \\
		\hline
		Predicted Negative & FN & TN \\
		\hline
		& P = TP+FN & N = FP+TN \\ 
		\hline
	\end{tabular}
\end{table}

\subsection{Accuracy}
It is one of the most commonly used evaluation metrics when it comes to interpreting the performance of ML algorithms. This is partly due to its' simplicity and ease of implement \cite{japkowicz2006question}. Accuracy can be defined as a measurement of how many test data points were correctly classified which is often shown in the form of a percentage. However, it is not recommended to carry out an accuracy measurement on the training data as if the algorithm is prone to overfitting as the accuracy rate can often be higher than what it actually is giving an unreliable result \cite{hall1999correlation}. Mathematically accuracy can be represented by:
\begin{equation}
Accuracy = \frac{TP+TN}{P+N}
\end{equation}
One of the biggest drawbacks with accuracy is its’ inability to differentiate between the FP and FN \cite{japkowicz2006question} therefore it is difficult to see where the algorithm is making errors which could lead to more serious complications depending on where the algorithm is being used. The computed accuracy values for binary and multiclass classifications are demonstrated in Table \ref{tab:Bin_class} and \ref{tab:Mul_class}, respectively.

\begin{table*}
	\center
	\setlength{\tabcolsep}{1em}
	\renewcommand{\arraystretch}{1.5}
	\caption{Results for binary classification.}
	\label{tab:Bin_class}       
	\begin{tabular}{llllll} 
		\hline
		Algorithm & Accuracy & Precision & Recall & F1 Score & Specificity  \\
		\hline                
		RF		  & 0.9977 	 & 	0.9632	 &    0.9644	  &	0.9638	 & 0.9988\\
		\hline
		SVM - Linear &0.9938 &	0.9955   &	  0.8380   & 0.9098	 & 0.9999\\
		\hline
		SVM - Quadratic &0.9954 & 0.9744 &	  0.8907   & 0.9307   & 0.9992\\
		\hline
		SVM - Medium Gaussian&0.9962 & 0.9423 & 0.9385 & 0.9404   & 0.9981\\ 
		\hline
		LR		  &    0.9937 & 0.9693   &	  0.8510	  & 0.8943	 & 0.9989\\
		\hline
	\end{tabular}
\end{table*}

\begin{table*}
	\center
	\setlength{\tabcolsep}{1em}
	\renewcommand{\arraystretch}{1.5}
	\caption{Results for multi classification.}
	\label{tab:Mul_class}       
	\begin{tabular}{llllll} 
		\hline
		Algorithm & Accuracy & Precision & Recall & F1 Score & Specificity  \\
		\hline                
		RF		  & 0.9294 	 & 	0.7493	 &    0.9991	  &	0.8564	 & 0.9963\\
		\hline
		SVM - Quadratic &0.9390 &	0.9753   &	  0.8771   & 0.9236 & 0.9869\\
		\hline
		SVM - Medium Gaussian&0.9338 & 1 &	  1   & 1  & 1\\
		\hline
		KNN - Fine&0.9982 & 0.9489 & 0.9570 & 0.9533& 0.9997\\ 
		\hline
		KNN - Coarse		  &    0.9934 & 0.8979   &	  0.9058	  & 0.9018	 & 0.9991\\
		\hline
		KNN - Weighted		  &    0.9983 & 0.9983   &	  0.9992	  & 0.9988	 & 0.9971\\
		\hline
	\end{tabular}
\end{table*}

\subsection{Precision and Recall}
Precision and recall are often joined together as they correlate with each other. Precision is defined as representing the number of positive predictions that are in the positive class. Recall is defined as the number of positive predictions out of all of the positive instances \cite{Brownlee2020}. The mathematical equations for both precision and recall are shown below. 
\begin{equation}
Precision = \frac{TP}{TP+FP}
\end{equation}
\begin{equation}
Recall = \frac{TP}{P}
\end{equation}
Similar to accuracy, precision and recall are commonly used due to their simplicity to implement and understand which is the main advantage of them. The main disadvantage with precision and recall is the lack of use of TN \cite{Doring2018}. This means that the negatives, that have been correctly classified, do not impact the overall score given to either of these metrics. This can therefore give an overall unbalanced view of the algorithms performance as TN are not considered and should only be considered when TN scores are not required.The computed precision and recall values for binary and multiclass classifications are demonstrated in Table \ref{tab:Bin_class} and \ref{tab:Mul_class}, respectively.

\subsection{F-Score}
Also known as F1 Score or F-Measure, it is the weighted average of both precision and recall which provides a single overall score for them. F-Score can be mathematically defined as:
\begin{equation}
F = \frac{2(Precision x Recall)}{(Precision + Recall)}
\end{equation}
It is considered a popular metric due to being able to give an overall view of both precision and recall as there are instances where an algorithm could have a high precision but low recall or vice versa \cite{Brownlee2020}. While this evaluation metric is widely used to give a more balanced view of precision and recall, F-Score as an evaluation metric should be used with caution \cite{hand2018note}. This is due to the metric taking the weighted average of precision and recall. The source states that the weighted value varies from calculation to calculation as the weighs are dependent on what is being evaluated. The authors in \cite{hand2018note} argue that to mitigate this issue there should be a standard weight which is used for all calculations of the metric.

\subsection{Specificity}
\begin{equation}
Specificity = \frac{TN}{N}
\end{equation}
This metric is often paired with sensitivity, which is also known as recall. As recall is being used for this project, sensitivity will not be used or mentioned. Due to recall/sensitivity dealing with TP and FN and specificity handling TN and FP, it can be argued that these two metrics together provide a more rounded overall accuracy compared to the actual accuracy metrics. This is due to accuracy not performing well with unbalanced datasets \cite{Doring2018}.

\subsection{Confusion Matrix}
\begin{figure}
	\centering
	\subfloat[]{\includegraphics[width=0.23\textwidth]{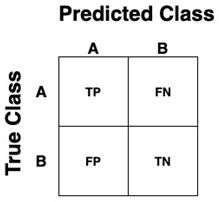} } \hspace*{0.2cm}
	\subfloat[]{\includegraphics[width=0.23\textwidth]{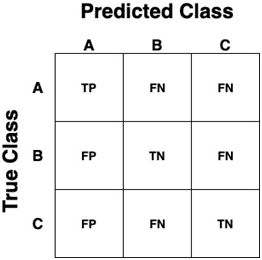} }
	\caption{Confusion matrices: (a) Binary classification confusion matrix, (b) Multiclass classification confusion matrix.}
	\label{Con_Matric}
\end{figure}
\begin{figure*}
	\centering
	\subfloat[]{\includegraphics[width=0.3\textwidth]{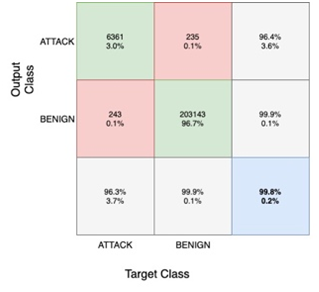} } 
	\hspace*{0.5cm}
	\subfloat[]{\includegraphics[width=0.3\textwidth]{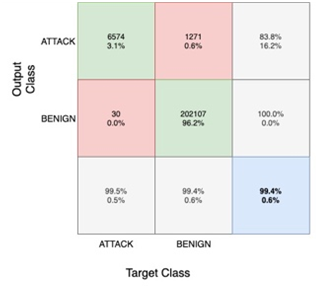} } 
	\hspace*{0.5cm}
	\subfloat[]{\includegraphics[width=0.3\textwidth]{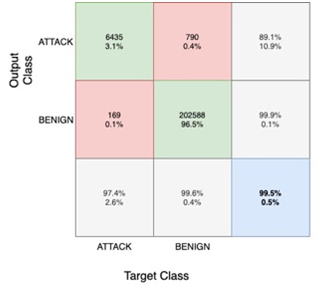} }   
	\\
	\vspace*{0.3cm}
	\subfloat[]{\includegraphics[width=0.3\textwidth]{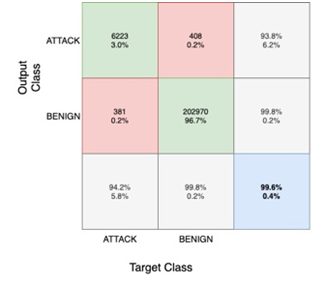} } 
	\hspace*{0.5cm}
	\subfloat[]{\includegraphics[width=0.3\textwidth]{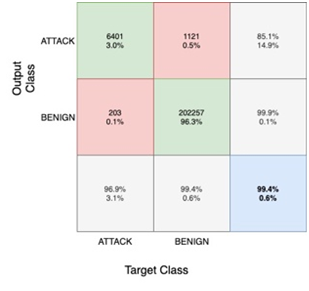} } 
	\hspace*{0.5cm}
	\subfloat[]{\includegraphics[width=0.3\textwidth]{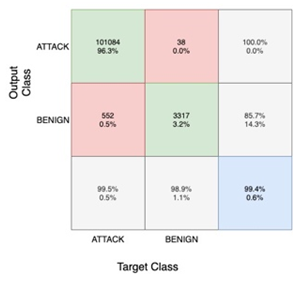} } 
	\\
	\vspace*{0.3cm}
	\subfloat[]{\includegraphics[width=0.3\textwidth]{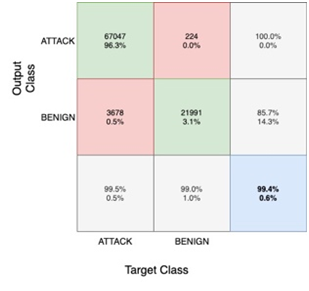} }
	\hspace*{0.5cm}
	\subfloat[]{\includegraphics[width=0.3\textwidth]{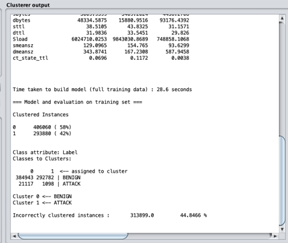} } 
	\caption{ Binary calculated confusion matrices: (a) random forest, (b) linear SVM, (c) quadratic SVM, (d) medium Gaussian SVM, (e) logistic regression, (f) NN test data, (g) NN all and (h) k-means clustering.}
	\label{Bin_Con_Matric}
\end{figure*}

The confusion matrix is an evaluation metric which take all four factors into account; TP, FP, FN, and TN \cite{ruuska2018evaluation}. This metric can be useful for visually understanding how well the algorithm performed and where the downfalls might be. For example, if a particular attack continually presented FN or FP readings then it could be a sign that the algorithm needs to be adjusted in order to be able to identify this attack \cite{Brownlee2016}. For a binary classification confusion matrix, a simple form is taken seen in Figure \ref{Con_Matric}(a). Whereas for a multiclass classification confusion matrix, this takes the form of \ref{Con_Matric}(b). From both the Figures \ref{Con_Matric}(a) and \ref{Con_Matric}(b) it can be seen where the factors involved are used and how the confusion matrix classifies each data point. This in turn makes it simpler to see where the algorithm is performing better and if it prone to FP or FN. Due to the confusion matrices taking into account all factors and having the ability to visually show results for each factor, this makes it a popular evaluation method particularly when trying to understand and improve an algorithms performance.\\

The results from the binary classification show that all the algorithms performed as expected in terms of accuracy with RF outperforming the other algorithms in terms of accuracy, precision, and F1 score with results of 0.9977, 0.9977 and 0.9989 respectively. \\

From the confusion matrices it can be seen in more detail that RF and medium Gaussian SVM both showed the same high perform level and displayed a similar trend in terms of correct and incorrect classifications. Though RF does show a slight decrease of 0.1$\%$ in FP and FN which make RF a better performer against medium Gaussian and the other algorithms in terms of binary classification. Despite being an overall better performer, RF does contain the second highest rating in terms of FN and while this seen as not as detrimental as a FP rating, it is still a high rating compared to the other algorithms such as linear SVM which shows 30 FN compared to RF’s 243. The computed confusion matrices for binary random forest, binary linear SVM, binary quadratic SVM, binary medium Gaussian SVM, binary logistic regression, binary NN and binary all are demonstrated in Figure \ref{Bin_Con_Matric}(a), (b), (c), (d), (e) and (f), respectively. \\

The screen shot from the Unsupervised algorithm implementation, Figure \ref{Bin_Con_Matric}(h), indicates that the k-means clustering algorithm did not perform well as it concluded with a 44.8$\%$ error rate which means nearly half of the data points were misclassified. This type of result shows that clustering algorithms are not suitable for this type of data as there was no distinct data that could form a solid centroid in order for the clusters to form. Though from  Figure \ref{Bin_Con_Matric}(h) it can be noted that the labels were successfully allocated to clusters although the data was incorrectly classified. It can also be noted where the algorithm made the wrong classifications. From clustered instances it can be noted that algorithms classified 42$\%$ of the dataset as an ATTACK meaning this algorithm was prone to a high number of FN so BENIGN traffic was being classed as an ATTACK when it should not have been. This could be put down to the features selected indicating that they were not suitable for this algorithm and did not provide enough information. \\

Overall all the supervised algorithms performed well in terms of the evaluation metrics with RF displaying a more promising performance compared to the other algorithms.\\

\begin{figure*}
	\centering
	\subfloat[Multiclass random forest confusion matrix.]{\includegraphics[width=0.3\textwidth]{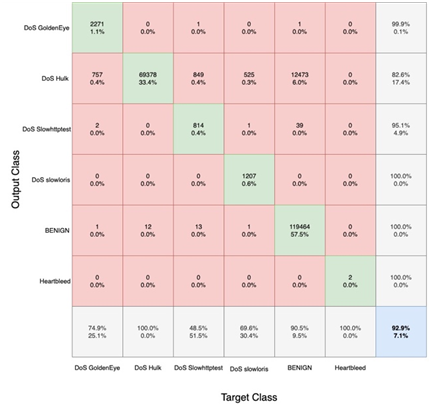} } 
	\hspace*{0.5cm}
	\subfloat[Multiclass quadratic SVM confusion matrix.]{\includegraphics[width=0.3\textwidth]{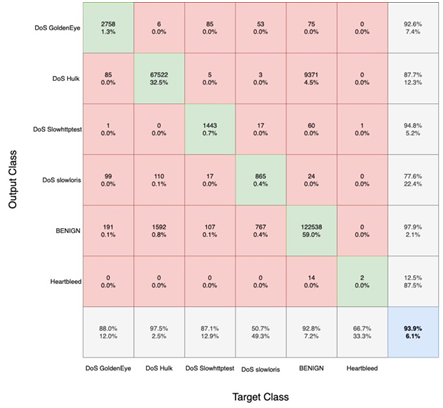} } 
	\hspace*{0.5cm}
	\subfloat[Multiclass fine KNN confusion matrix.]{\includegraphics[width=0.3\textwidth]{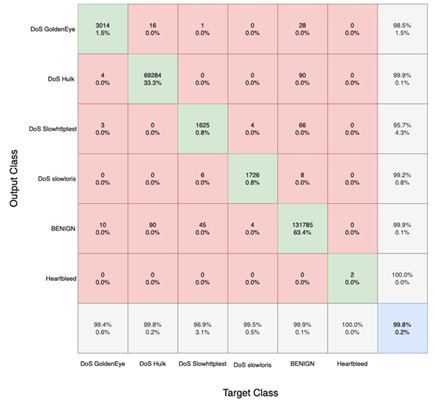} }   
	\\
	\vspace*{0.3cm}
	\subfloat[Multiclass coarse KNN confusion matrix.]{\includegraphics[width=0.3\textwidth]{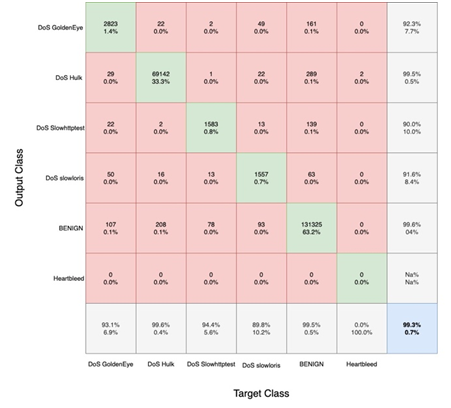} } 
	\hspace*{0.5cm}
	\subfloat[Multiclass weighted KNN confusion matrix.]{\includegraphics[width=0.3\textwidth]{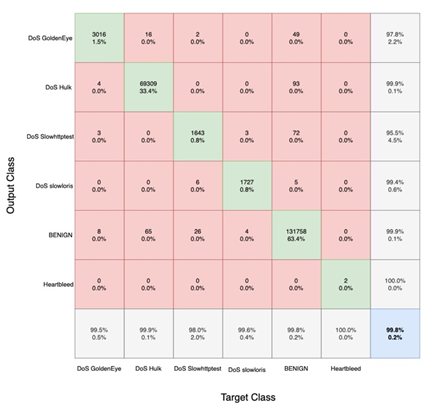} } 
	\hspace*{0.5cm}
	\subfloat[Multiclass medium Gaussian SVM confusion matrix.]{\includegraphics[width=0.3\textwidth]{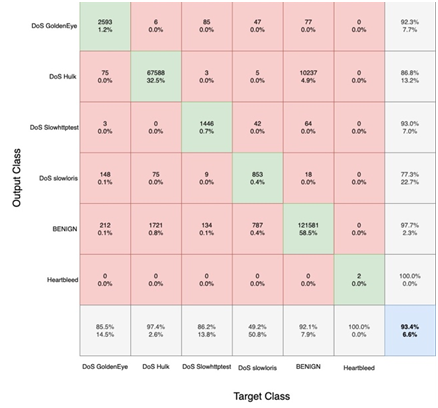} } 
	\\
	\vspace*{0.3cm}
	\subfloat[Multiclass NN test confusion matrix.]{\includegraphics[width=0.3\textwidth]{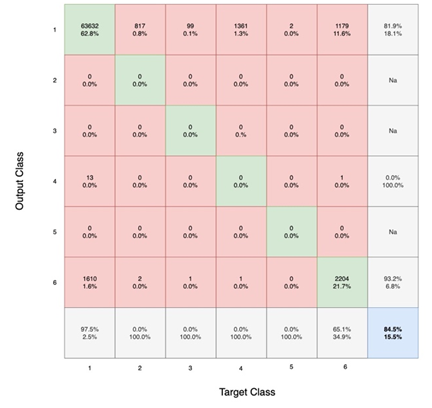} } 
	\hspace*{0.5cm}
	\subfloat[Multiclass NN all confusion matrix.]{\includegraphics[width=0.3\textwidth]{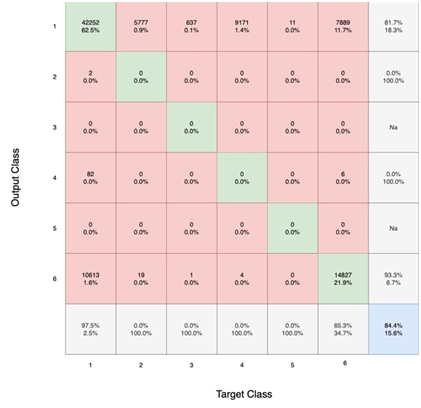} } 
	\hspace*{0.5cm}
	\subfloat[Multiclass k-means clustering results.]{\includegraphics[width=0.3\textwidth]{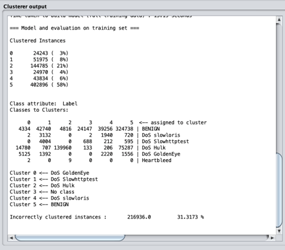} } 
	\caption{Multiclass calculated confusion matrices.}
	\label{Multi_Con_Matric}
\end{figure*}

The computed confusion matrices for multiclass confusion are shown in Figure \ref{Multi_Con_Matric}. The results for the multiclass classification show quite high rankings in terms of accuracy with all the algorithms performing well. From a more in-depth analysis of the results that it is evident that the KNN algorithms greatly outperformed the other algorithms in terms of accuracy, precision and F1 Score. The greater performers out of the KNN algorithms were the Fine KNN and Weighted KNN algorithms, both of which possessed an F1 Score of 0.9991. However Weighed just narrowly outperformed Fine in terms of accuracy and precision with a score of 0.9983 compared to 0.9982 respectively. \\

In terms of confusion matrices, it can be visually seen that the KNN algorithms had significantly lower FN and FP rates compared to the other algorithms. For example, between medium Gaussian SVM and Fine KNN it can be noted that there is a substantial difference when predicting a “DoS slowloris” attack had occurred. With medium Gaussian SVM it can be seen that 50.8$\%$ of these were misclassified compared to Fine KNN where 0.5$\%$ of these were misclassified. It can also be noted that NN did not perform well with the multiclass classification with very poor TP rates which can be seen in 4 out of the 6 classes.
The results from the unsupervised algorithm implementation show that the k-means Clustering algorithm did not perform well as it concluded with a 31.3$\%$ error rate meaning nearly a third of the data points were misclassified. This type of result shows that clustering algorithms are not suitable for this type of data. While five of the labels were successfully assigned to clusters Heartbleed was not allocated, this could be due to that fact there was not a great number of Heartbleed attacks within this dataset. \\
While algorithms such as SVM, RF, and NN did not perform poorly as an overall accuracy, it can be seen through the confusion matrices that these algorithms were more prone to FN and FP making them less effective when considering intrusion detection, especially when compared to KNN algorithms.

\section{\textbf{Comparison}} 
All the results above prove that ML can be successful and of use when it comes to detecting whether an attack has occurred or not when given particular characteristics such as source IP, destination IP, and packet length etc. Between supervised and unsupervised learning, it is clear that the supervised learning is more successful and effective though only one unsupervised method has been looked at, so it is not as easy to give a full comparison between the two methods. \\
\\
In terms of identifying whether the algorithms are more effective in binary classification or multiclass classification, the binary classification results ate more evident of an overall high performance-level across all the algorithms. This kind of result demonstrate that in order to simply determine whether an event has taken place or not, it can be effectively achieved using ML algorithms. The results also demonstrated that for binary classification, RF is most effective in terms of high accuracy, high precision, and a high F1 Score. This algorithm also prove to be effective in terms of confusion matrix, displaying low scores in FN’s, and FP’s therefore increasing its effectiveness and reliability of thd algorithm.\\

\begin{figure}
	\centering
	\subfloat[Comparison of accuracy scores.]{\includegraphics[width=0.5\textwidth]{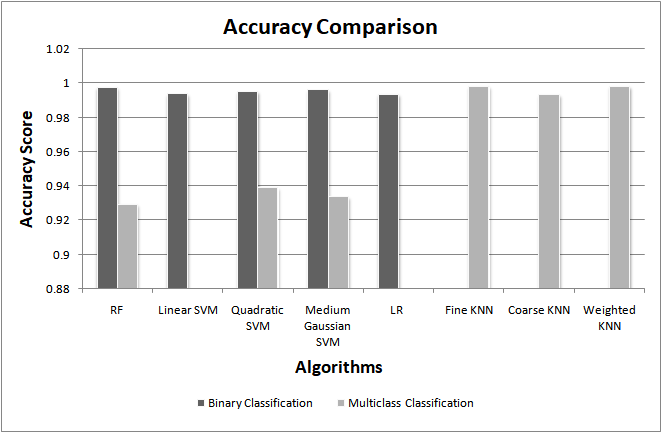} }   
	\\
	\vspace*{0.3cm}
	\subfloat[Comparison of precision scores.]{\includegraphics[width=0.5\textwidth]{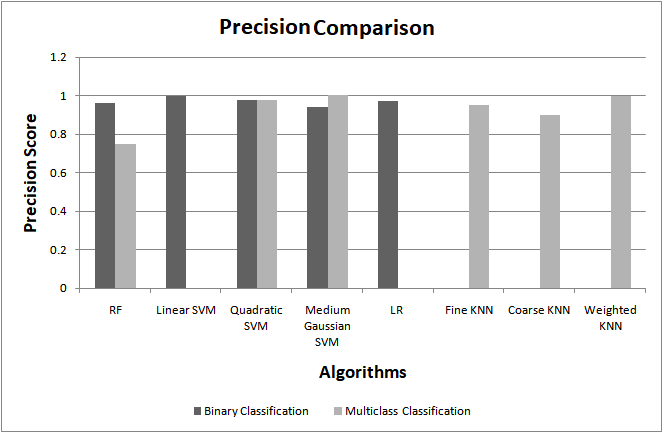} } 
	\\
	\vspace*{0.3cm}
	\subfloat[Comparison of F1-scores.]{\includegraphics[width=0.5\textwidth]{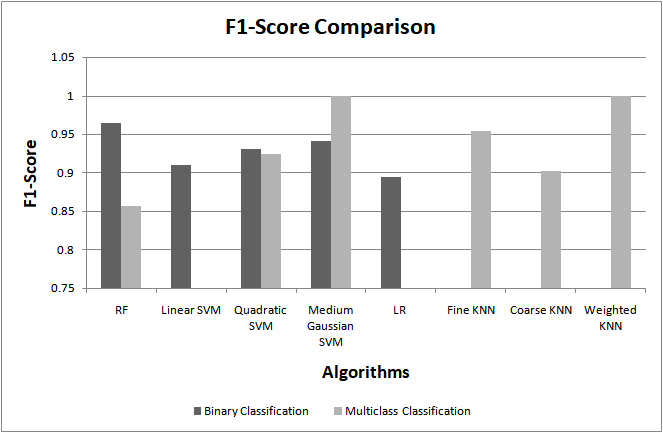} } 
	\caption{Binary and multiclass classification comparison.}
	\label{comp}
\end{figure}

For the multiclass classification results, it could be seen in Figure \ref{comp} that there was a more spread out range of results. Though all the results are still relatively high, there is a more distinct difference in the performance levels of the different algorithms. While RF performs well within binary classification, the results shown for multiclass classification rank RF as the least effective algorithm in terms of accuracy, precision, and F1 Score with both types of SVM not far ahead. For the multiclass it is evident that the KNN algorithms is most effective in terms of all the evaluation metrics. Multiclass classification is considered more complex than binary classification due to the nature of the classification being more than two classes; this in turn causes strains on algorithms through computational power and time and can therefore lead to less effective result from the algorithm as seen from the results throughout the experiment within this work.\\

From Figures. \ref{comp}(a), (b), and (c), a trend can be noted that algorithms that performed well in binary classification did not perform as well when it came to multiclass classification. This difference is particularly noticeable within the RF accuracy scores where there is a 7$\%$ difference between binary and multiclass classification. A similar score difference can be seen for RF in precision and F1 -score. Despite this, it can also be noted that KNN did not only outperform the other algorithms used within the multiclass classification by 6.61$\%$, when compared with quadratic SVM, but also the algorithms within the binary classification by 0.06$\%$ when compared with RF. This in turn shows that KNN are strong algorithms when it comes to classification of data points and has proved to be the most effective algorithm.\\

While the unsupervised algorithm did not perform well it is important to note that the multiclass classification did perform better with a greater number of data points, 31.3$\%$ being correctly assigned to a cluster compared to binary with 44.8$\%$. This does illustrate that the algorithm is more suitable for multiclass classification and the result could potentially be improved by using a different dataset that is balanced.\\

Overall between the two types of classification, it can be seen that binary classification does produce higher results when considering accuracy, precision, F1 Score, and confusion matrices, as this type of classification is simpler to implement and is primarily what the algorithms are designed for. Given the results produced in this work, it is possible to conclude that all the algorithms used for binary classification would be suitable to use in an intrusion detection system of that type. However, regarding multiclass classification it would not be sensible to recommend all the algorithms due to the large number of FN’s and FP’s that occur with a number of the algorithms. Instead only KNN algorithms can be recommended with Weighted KNN being the most obvious choice.

\section{Conclusion}
This paper presents an extensive overview, implementation, and cross comparison of state of the art machine learning based methods available for intrusion detection. It is evident from the investigations that different machine learning methods can be used for intrusion detection. Further, the results demonstrated that the usage of machine learning techniques produce positive impact on improving the overall performance of the intrusion detection system in terms of accuracy and lowering false negatives. Additionally, from the perspective of algorithms, KNN showed the most promising results when considering the metrics of accuracy, recall, F1-score, and confusion with giving near-perfect results. In addition to providing the best results, it was also noted that multiclass classification can provide more informative results by differentiating between different types of attack, therefore, allowing for a more effective response when mitigating the impact of such an attack.\\
As a future works, we plan to investigate the use of deep learning methods for intrusion detection \cite{hajjaji2021big,boulila2021rs,boulila2021novel}. 

\bibliographystyle{IEEEtran}
\bibliography{Ref}

\begin{thebibliography}{10}
\providecommand{\url}[1]{#1}
\csname url@samestyle\endcsname
\providecommand{\newblock}{\relax}
\providecommand{\bibinfo}[2]{#2}
\providecommand{\BIBentrySTDinterwordspacing}{\spaceskip=0pt\relax}
\providecommand{\BIBentryALTinterwordstretchfactor}{4}
\providecommand{\BIBentryALTinterwordspacing}{\spaceskip=\fontdimen2\font plus
\BIBentryALTinterwordstretchfactor\fontdimen3\font minus
  \fontdimen4\font\relax}
\providecommand{\BIBforeignlanguage}[2]{{%
\expandafter\ifx\csname l@#1\endcsname\relax
\typeout{** WARNING: IEEEtran.bst: No hyphenation pattern has been}%
\typeout{** loaded for the language `#1'. Using the pattern for}%
\typeout{** the default language instead.}%
\else
\language=\csname l@#1\endcsname
\fi
#2}}
\providecommand{\BIBdecl}{\relax}
\BIBdecl

\bibitem{bhuyan2013network}
M.~H. Bhuyan, D.~K. Bhattacharyya, and J.~K. Kalita, ``Network anomaly
  detection: methods, systems and tools,'' \emph{Ieee communications surveys \&
  tutorials}, vol.~16, no.~1, pp. 303--336, 2013.

\bibitem{finnerty2019cyber}
K.~Finnerty, S.~Fullick, H.~Motha, J.~N. Shah, M.~Button, and V.~Wang, ``Cyber
  security breaches survey 2019,'' 2019.

\bibitem{huma2021hybrid}
Z.~E. Huma, S.~Latif, J.~Ahmad, Z.~Idrees, A.~Ibrar, Z.~Zou, F.~Alqahtani, and
  F.~Baothman, ``A hybrid deep random neural network for cyberattack detection
  in the industrial internet of things,'' \emph{IEEE Access}, vol.~9, pp.
  55\,595--55\,605, 2021.

\bibitem{denning1987intrusion}
D.~E. Denning, ``An intrusion-detection model,'' \emph{IEEE Transactions on
  software engineering}, no.~2, pp. 222--232, 1987.

\bibitem{liu2014intrusion}
G.~G. Liu, ``Intrusion detection systems,'' in \emph{Applied Mechanics and
  Materials}, vol. 596.\hskip 1em plus 0.5em minus 0.4em\relax Trans Tech Publ,
  2014, pp. 852--855.

\bibitem{chio2018machine}
C.~Chio and D.~Freeman, \emph{Machine Learning and Security: Protecting Systems
  with Data and Algorithms}.\hskip 1em plus 0.5em minus 0.4em\relax " O'Reilly
  Media, Inc.", 2018.

\bibitem{ali2020network}
A.~Ali, S.~Shaukat, M.~Tayyab, M.~A. Khan, J.~S. Khan, J.~Ahmad \emph{et~al.},
  ``Network intrusion detection leveraging machine learning and feature
  selection,'' in \emph{2020 IEEE 17th International Conference on Smart
  Communities: Improving Quality of Life Using ICT, IoT and AI (HONET)}.\hskip
  1em plus 0.5em minus 0.4em\relax IEEE, 2020, pp. 49--53.

\bibitem{shaukat2020intrusion}
S.~Shaukat, A.~Ali, A.~Batool, F.~Alqahtani, J.~S. Khan, J.~Ahmad
  \emph{et~al.}, ``Intrusion detection and attack classification leveraging
  machine learning technique,'' in \emph{2020 14th International Conference on
  Innovations in Information Technology (IIT)}.\hskip 1em plus 0.5em minus
  0.4em\relax IEEE, 2020, pp. 198--202.

\bibitem{khan2021iot}
M.~A. khan, M.~A. Khan, S.~Latif, A.~A. Shah, M.~U. Rehman, W.~Boulila,
  M.~Driss, and J.~Ahmad, ``Voting classifier-based intrusion detection for iot
  networks,'' in \emph{2021 2nd International Conference of Advance Computing
  and Informatics (ICACIN)}.\hskip 1em plus 0.5em minus 0.4em\relax Springer,
  2021.

\bibitem{tidjon2019intrusion}
L.~N. Tidjon, M.~Frappier, and A.~Mammar, ``Intrusion detection systems: A
  cross-domain overview,'' \emph{IEEE Communications Surveys \& Tutorials},
  vol.~21, no.~4, pp. 3639--3681, 2019.

\bibitem{shenfield2018intelligent}
A.~Shenfield, D.~Day, and A.~Ayesh, ``Intelligent intrusion detection systems
  using artificial neural networks,'' \emph{ICT Express}, vol.~4, no.~2, pp.
  95--99, 2018.

\bibitem{rege2018machine}
M.~Rege and R.~B.~K. Mbah, ``Machine learning for cyber defense and attack,''
  \emph{DATA ANALYTICS 2018}, p.~83, 2018.

\bibitem{latif2020novel}
S.~Latif, Z.~Zou, Z.~Idrees, and J.~Ahmad, ``A novel attack detection scheme
  for the industrial internet of things using a lightweight random neural
  network,'' \emph{IEEE Access}, vol.~8, pp. 89\,337--89\,350, 2020.

\bibitem{Keshari2020}
K.~Keshari, \emph{Top 10 Applications of Machine Learning: Machine Leaning
  Applications in Daily Life}, 2020.

\bibitem{Sober2020}
R.~Sober, \emph{Data Breach Response Times: Trends and Tips}, 2020.

\bibitem{shafique2021detecting}
A.~Shafique, J.~Ahmed, W.~Boulila, H.~Ghandorh, J.~Ahmad, and M.~U. Rehman,
  ``Detecting the security level of various cryptosystems using machine
  learning models,'' \emph{algorithms}, vol.~1, p.~5, 2021.

\bibitem{amin2017accelerated}
M.~S. Amin, L.~Hassan, A.~A. Shah, U.~Akbar, and H.~A. Niaz, ``Accelerated gpu
  based protein sequence alignment--an optimized database sequences approach,''
  \emph{IJCSNS}, vol.~17, no.~10, p. 231, 2017.

\bibitem{Wiggers2019}
K.~Wiggers, \emph{Gmail is now blocking 100 million more spam emails a day,
  thanks to TensorFlow}, 2019.

\bibitem{japkowicz2006question}
N.~Japkowicz, ``Why question machine learning evaluation methods,'' in
  \emph{AAAI workshop on evaluation methods for machine learning}, 2006, pp.
  6--11.

\bibitem{hall1999correlation}
M.~A. Hall, ``Correlation-based feature selection for machine learning,'' 1999.

\bibitem{Brownlee2020}
J.~Brownlee, \emph{How to Calculate Precision, Recall, and F-Measure for
  Imbalanced Classification}, 2020.

\bibitem{Doring2018}
M.~Döring, \emph{The Case Against Precision as a Model Selection Criterion},
  2018.

\bibitem{hand2018note}
D.~Hand and P.~Christen, ``A note on using the f-measure for evaluating record
  linkage algorithms,'' \emph{Statistics and Computing}, vol.~28, no.~3, pp.
  539--547, 2018.

\bibitem{ruuska2018evaluation}
S.~Ruuska, W.~H{\"a}m{\"a}l{\"a}inen, S.~Kajava, M.~Mughal, P.~Matilainen, and
  J.~Mononen, ``Evaluation of the confusion matrix method in the validation of
  an automated system for measuring feeding behaviour of cattle,''
  \emph{Behavioural processes}, vol. 148, pp. 56--62, 2018.

\bibitem{Brownlee2016}
J.~Brownlee, \emph{What is a Confusion Matric in Machine Learning}, 2016.

\bibitem{hajjaji2021big}
Y.~Hajjaji, W.~Boulila, I.~R. Farah, I.~Romdhani, and A.~Hussain, ``Big data
  and iot-based applications in smart environments: A systematic review,''
  \emph{Computer Science Review}, vol.~39, p. 100318, 2021.

\bibitem{boulila2021rs}
W.~Boulila, M.~Sellami, M.~Driss, M.~Al-Sarem, M.~Safaei, and F.~A. Ghaleb,
  ``Rs-dcnn: A novel distributed convolutional-neural-networks based-approach
  for big remote-sensing image classification,'' \emph{Computers and
  Electronics in Agriculture}, vol. 182, p. 106014, 2021.

\bibitem{boulila2021novel}
W.~Boulila, H.~Ghandorh, M.~A. Khan, F.~Ahmed, and J.~Ahmad, ``A novel
  cnn-lstm-based approach to predict urban expansion,'' \emph{Ecological
  Informatics}, p. 101325, 2021.

\end{thebibliography}

\end{document}